# Perform wordcount Map-Reduce Job in Single Node Apache Hadoop cluster and compress data using Lempel-Ziv-Oberhumer (LZO) algorithm

Nandan Mirajkar[1], Sandeep Bhujbal[2], Aaradhana Deshmukh[3]

[1]Department of Advanced Software and Computing Technologies
IGNOU – I²IT Centre of Excellence for Advanced Education and Research
Pune, Maharashtra 411 057, India

[2]Department of Advanced Software and Computing Technologies
IGNOU – I²IT Centre of Excellence for Advanced Education and Research
Pune, Maharashtra 411 057, India

[3]Department of Computer Engineering
Smt. Kashibai Navale College of Engineering. Vadgaon (Bk), Off Sinhgad Road.
Pune, Maharashtra 411 041, India

## Abstract

Applications like Yahoo, Facebook, Twitter have huge data which has to be stored and retrieved as per client access. This huge data storage requires huge database leading to increase in physical storage and becomes complex for analysis required in business growth. This storage capacity can be reduced and distributed processing of huge data can be done using Apache Hadoop which uses Map-reduce algorithm and combines the repeating data so that entire data is stored in reduced format. The paper describes performing a wordcount Map-Reduce Job in Single Node Apache Hadoop cluster and compress data using Lempel-Ziv-Oberhumer (LZO) algorithm.

## 1.0 Introduction

Hadoop was created by Doug Cutting an employee at Yahoo and Michael J. Cafarella. It was originally developed to support distribution for the Nutch search engine project [12]. Hadoop was inspired by papers published by Google regarding its approach in handling an avalanche of data, and became a standard for storing, processing and analyzing hundreds of terabytes, and even petabytes of data. Hadoop's breakthrough advantages mean that businesses and organizations can now find value in data that was recently considered useless [11]. Hadoop enables a computing solution that is: Cost effective – Due to massive parallel computing approach by Hadoop, there is decrease in the cost per terabyte of storage. Fault tolerant – When a node is missed or a fault arises the system navigates work to another location of the data and continues processing. Flexible – Hadoop is schema-less, and can accept any type of data,structured or not, from any number of sources. Data from multiple sources can be joined and aggregated in arbitrary ways enabling deeper analyses than any one system can provide. Scalable – New nodes can be added as required and added without changing data formats [13].

Compression reduces number of bytes read from or written to HDFS. Compression enhances efficiency of network bandwidth and disk space. HBase is used when need arises for random, realtime read/write access to Big Data [19]. HBase comes with only Gzip compression, compression by GZip is not as fast as Lempel-Ziv-Oberhumer (LZO) compression. For maximum performance LZO is used but HBase cannot ship with LZO because of the licensing issues hence LZO installation is done post-HBase installation [18]. The section 2 of this paper depicts information related to Apache Hadoop , Hadoop distributed file system (HDFS) and its architecture, Map-reduce technique. Section 3 gives details about installation of single node Hadoop cluster on Ubuntu 10.04 server edition open source operating system. Section 3 also gives results of wordcount example using Hadoop which is a Map-reduce job. Section 4 describes installation of Hbase and its usage in Hadoop cluster. Section 5 gives an example of compression of data in Hadoop using Lempel-Ziv-Oberhumer (LZO) algorithm.

## 2.0 Hadoop

### 2.1 Apache Hadoop

The framework Apache Hadoop is used for distributed processing of huge data sets known as "Big Data" across clusters of computers using a simple programming model [2][1]. Hadoop permits an application to map, group, and reduce data across a distributed cloud of machines so that applications can process huge data [1]. It can scale up to large number of machines as required for the job; each machine will provide local computation and storage. Apache Hadoop software library itself detects and handles any failures at application layer [2].

## 2.2 Hadoop Distributed File System - HDFS

A distributed user-level filesystem HDFS─Hadoop Distributed File System written in Java [15] stores huge files across machines in a large cluster. Hadoop DFS stores each file as a sequence of blocks, all blocks in a file except the last block are the same size typically 64 MB [14][15]. Blocks belonging to a file are replicated for fault tolerance. The block size and replication factor are configurable per file. An application can specify the number of replicas of a file. The replication factor can be specified at file creation time and can be changed later. Files in HDFS are "write once" and have strictly one writer at any time [15][16].

## 2.3 Architecture

HDFS comprises of interconnected clusters of nodes where files and directories reside. An HDFS cluster consists of a single node, known as a Name-Node that manages the file system namespace and regulates client access to files. In addition, Data-Nodes store data as blocks within files [27].

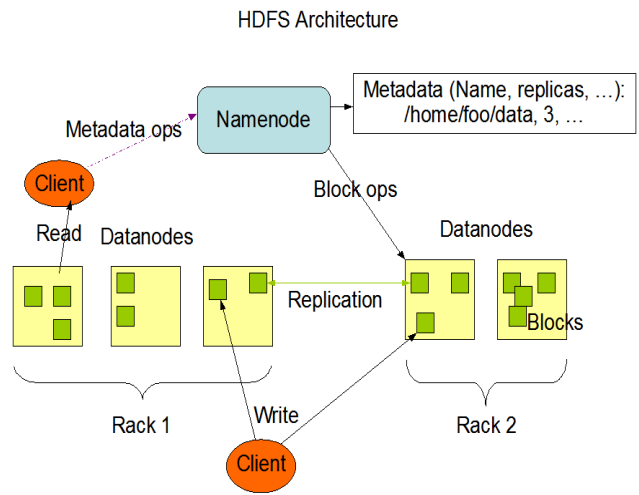

Fig 1: HDFS Architecture [17]

## 2.4 Name-Node

The Name-Node executes file system namespace operations like opening, closing, and renaming files and directories. The Name-Node does not store HDFS data itself, but rather maintains a mapping between HDFS file name, a list of blocks in the file, and the Data Node on which those blocks are stored. The Name-Node makes all decisions regarding replication of blocks [15][17].

## 2.5 Secondary Name-Node

HDFS includes a Secondary Name-Node, there is a misconception that secondary Name-Node comes into action after Primary Name-Node (i.e Name-Node) fails. The fact is Secondary Name-Node is continuously connected with Primary Name-Node and takes snapshots of Name-Node's memory structures. These snapshots can be used to recover the failed Name-Node and recent memory structure [12].

## 2.6 Data-Node

A Data-Node stores data in the Hadoop File System. A functional filesystem has more than one Data-Node, with data replicated across them. On startup, a Data-Node connects to the Name-Node; spinning until that service comes up. It then responds to requests from the Name-Node for filesystem operations. Client applications can talk directly to a Data-Node, once the Name-Node has provided the location of the data [22].

## 2.7 Job-Tracker

Job-Tracker keeps track of which Map-Reduce jobs are executing, schedules individual Maps, Reduces or intermediate merging operations to specific machines, monitors the success and failures of these individual Tasks, and works to complete the entire batch job. The Job-Tracker is a point of failure for the Hadoop Map-Reduce service. If it goes down, all running jobs are halted [20].

## 2.8 Task-Tracker

A Task-Tracker is a node in the Hadoop cluster that accepts tasks such as Map, Reduce and Shuffle operations from a Job-Tracker. Task-Tracker is set up with set of slots which depicts the number of tasks it can accept. The Task-Tracker spawns a separate JVM processes to do the actual work. The Task-Tracker supervises these spawned processes, capturing the output and exit codes. When the process finishes, successfully or not, the task tracker notifies the Job-Tracker. The Task-Trackers also transmit heartbeat messages to the Job-Tracker, usually every few minutes, to reassure the Job-Tracker that it is still alive. These messages also inform the Job-Tracker of the number of available slots, so the Job-Tracker can stay up to date with where in the cluster work can be assigned [20].

## 2.9 Map-reduce

Map-reduce is a Parallel Programming approach used for extracting and analyzing information from unstructured Big Data storage [8]. In Map-reduce their is a map function that processes a key/value pair to generate a set of intermediate key/value pairs, and a reduce function which will immix all intermediate values associated with the same intermediate key[7].

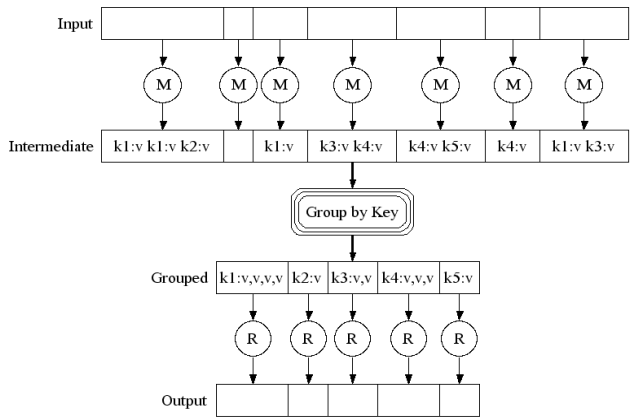

Fig 2: Map-reduce structure [28]

Mapping (M) is done on input data to get intermediate key/value pairs as shown in Figure 2 then this intermediate data is grouped by key e.g all values v with key k1 in one group , all values v with key k2 etc. This grouped data is reduced to give following output i.e. {k1, 4} {k2, 1} {k3, 2} {k4, 3} and {k5, 1}

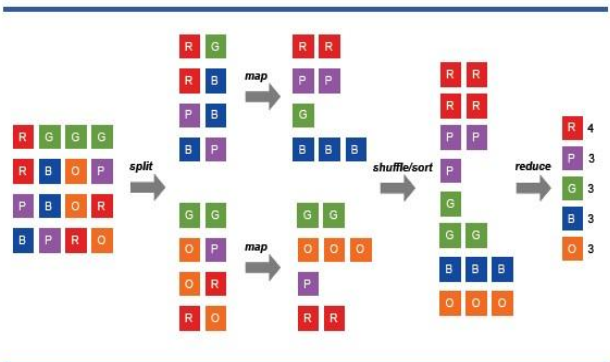

Fig 3: Map reduce example [21]

Above is example of color data on which map-reduce is performed. The 16 blocks are split into two sets with 8 blocks in each and it is mapped to arrange blocks with respect to colors where R is red, P is pink, G is green, B is blue and O is orange. The first set contains 2 red, 2 pink, 1 green, 3 blue and second set contains 2 green, 3 orange, 1 pink, 2 red. These 2 sets are shuffled or sorted to get a single set and is reduced to give following output {R, 4} {P, 3} {G, 3} {B, 3} {O, 3}.

Following is a code for map-reduce: The mapper read-s input records and produces <word, 1> as intermediate pairs. After shuffling, intermediate counts associated with the same word are passed to a reducer, which adds the counts together to produce the sum [8].

```
map(String key, String value)
    {
        for(each word w in value)
        {
        EmitIntermediate(w, 1);
        }
    }
```

In above case key: document name and value: document contents.

```
reduce(String key, Iterator values)
    {
        int sum = 0;
        for(each v in values)
        {
        sum += v;
        }
        Emit(key, sum);
    }
```

In above case key: a word and values: a list of counts.

## 3.0 Installation of Hadoop

### 3.1 Install Ubuntu 10.04 server

1.Insert Ubuntu 10.04 server edition and select "Install ubuntu server"
2.Select "configure network manually"
IP address: 192.168.0.216
Netmask: 255.255.255.0
Gateway: 192.168.0.1
Nameserver: 202.138.xx.x
3. Hostname give as 'n1'
4. Fullname for the new user: nandan
5. Software selection: select only "Open SSH server"
6. Install grub boot loader to master boot loader:yes
7. Complete installation and reboot the system
8. Now update the system using following command
*sudo apt-get update*
9. If desktop is required following command can be used
*sudo apt-get install ubuntu-desktop*

### 3.2 Install SUN-Java

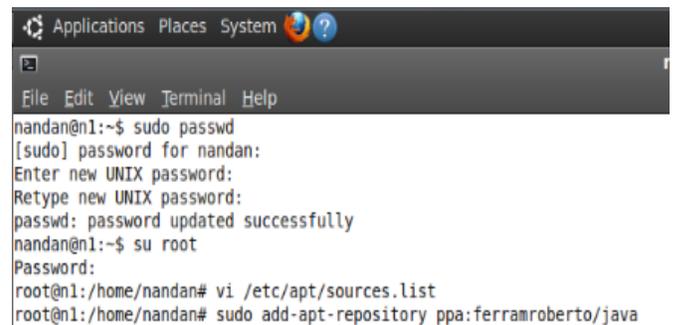

Fig 4: Screenshot of password update and installation of PPA

Login as user "nandan" and change the password of root. For Ubuntu 10.04 LTS, the sun-java6 packages have been dropped from the Multiverse section of the Ubuntu archive. It is recommended to use openjdk-6 instead. If one cannot switch from the proprietary Sun JDK/JRE to OpenJDK, install sun-java6 packages from the Canonical Partner Repository [29]. Any of the repositories can be added or modified by directly editing the files in /etc/apt/sources.list [30] so login as root and open sources.list file and add following lines in it at the end,

*deb http://archive.canonical.com/ lucid partner*

The JDK has tools needed for developing and testing programs written in the Java programming language and running on the Java Platform [5]. Sun-java6-jdk has been removed from the default Ubuntu 10.04 hence repositories are required, add a PPA (Personal Package Archives) repository then update the source list and install java6-jdk [3].Personal Package Archives (PPA) allows to upload Ubuntu source packages to be built and published as an apt repository by Launchpad [4]. Accept the operating system distributor license for Java during installation.

*root@n1:/home/nandan# sudo apt-get update*
*root@n1:/home/nandan# sudo apt-get install sun-java6-jdk*
*root@n1:/home/nandan# java -version*

## 3.3 Create Hadoop group and user

Add a group hadoop and user hadoop in same group, choose default for user information for hadoop [31]

*root@n1:/home/nandan# addgroup hadoop*
*root@n1:/home/nandan# adduser --ingroup hadoop hadoop*

Now Configuring sudo to allow users in the group "hadoop" to run commands as root [35].

*root@n1:/home/nandan# visudo*
root   ALL=(ALL) ALL
hadoop ALL=(ALL) ALL

To manage nodes Hadoop requires SSH access, remote machines plus local machine if one want's to use Hadoop on it. For single node setup configure SSH access to localhost for the hadoop user [6].
*root@n1:/home/nandan# su - hadoop*
*hadoop@n1:~$ ssh-keygen -t rsa -P ""*

The above command will create an RSA key pair with an empty password. Empty password is not recommended, but in this case it is needed to unlock the key so that one doesn't require entering the passphrase every time Hadoop interacts with its nodes [6].

Enable SSH access to local machine with this newly created key.

*hadoop@n1:~$   cat   /home/hadoop/.ssh/id_rsa.pub   >>*
*/home/hadoop/.ssh/authorized_keys*

Test the SSH setup by connecting to local machine with the hadoop user here local machine's host key fingerprint is saved to hadoop user's known_hosts file [6]. Then exit localhost.

*hadoop@n1:~$ ssh localhost*
*hadoop@n1:~$ exit*
*logout*
*Connection to localhost closed.*

## 3.4 Install Hadoop software

Download Hadoop from the Apache Download Mirrors and extract the contents of hadoop in /usr/local. Change the owner of all files to hadoop user and hadoop group using *chown* command [6].

nandan@n1:~$ su - root
Password: *****
root@n1:~# cd /usr/local/
root@n1:/usr/local#wget
http://apache.communilink.net/hadoop/core/hadoop-0.20.2/hadoop-0.20.2.tar.gz
root@n1:/usr/local# tar -xvf hadoop-0.20.2.tar.gz
root@n1:/usr/local# chown -R hadoop:hadoop hadoop-0.20.2
root@n1:/usr/local# ln -s hadoop-0.20.2/ hadoop
ln command lets a file/directory on disk be accessed with more than one file/directory name, hadoop can be used instead of hadoop-0.20.2/

Remove the tar file after extraction
*root@n1:/usr/local# rm -rf hadoop-0.20.2.tar.gz*

## 3.5 Configure the Hadoop

One problem with IPv6 on Ubuntu is that using 0.0.0.0 for the various networking-related Hadoop configuration options will result in Hadoop binding to the IPv6 addresses of Ubuntu box. One can disable IPv6 only for Hadoop by adding the lines of IPv4 shown in figure to *conf/hadoop-env.sh*. The environment variable that has to be configured for Hadoop is JAVA_HOME. Open *conf/hadoop-env.sh* set the JAVA_HOME environment variable to the Sun JDK/JRE 6 directory

*hadoop@n1:/usr/local/hadoop$ vi conf/hadoop-env.sh*

export JAVA_HOME=/usr/lib/jvm/java-6-sun
export HADOOP_OPTS="-Djava.net.preferIPv4Stack=true"

As of Hadoop 0.20.x and 1.x, the configuration settings previously found in hadoop-site.xml were moved to core-site.xml (hadoop.tmp.dir, fs.default.name), mapred-site.xml (mapred.job.tracker) and hdfs-site.xml (dfs.replication).

Configure the directory where Hadoop will store its data files, the network ports it listens to, etc. The hadoop.tmp.dir variable can be changed to the directory of own choice. Here the directory is /home/hadoop/cloud. Hadoop's default configurations use hadoop.tmp.dir as the base temporary directory both for the local file system and HDFS [6].If required localhost can be replaced with n1 in following xml files.

*hadoop@n1:/usr/local/hadoop$ mkdir ~/cloud*
*hadoop@n1:/usr/local/hadoop$ vi conf/core-site.xml*

```
<configuration>
<property>
<name>hadoop.tmp.dir</name>
<value>/home/hadoop/cloud/hadoop-${user.name}</value>
</property>
<property>
<name>fs.default.name</name>
<value>hdfs://localhost:9000</value>
<description>The name of the default file system.
</description>
</property>
</configuration>
```

*hadoop@n1:/usr/local/hadoop$ vi conf/mapred-site.xml*
```
<configuration>
<property>
<name>mapred.job.tracker</name>
<value>localhost:9001</value>
<description>The host and port that the MapReduce job tracker runs at. If 'local', then jobs are run in-process as a single map and reduce task.
</description>
</property>
</configuration>
```

*hadoop@n1:/usr/local/hadoop$ vi conf/hdfs-site.xml*
```
<configuration>
<property>
<name>dfs.replication</name>
<value>1</value>
<description>Default block replication.
The actual number of replications can be specified when the file is created. The default is used if replication is not specified in create time.
</description>
</property>
</configuration>
```

## 3.6 The Word-count Map-reduce Job

As it is single node Hadoop cluster with master and slave on same machine, n1 should be mentioned in master and slave files.

*hadoop@n1:/usr/local/hadoop$ vi conf/masters*
**n1**
*hadoop@n1:/usr/local/hadoop$ vi conf/slaves*
**n1**

The first step in starting up Hadoop installation is formatting the Hadoop filesystem, which is implemented on top of the local filesystems of cluster. This has to be carried out when first time Hadoop installation is done. Do not format a running Hadoop filesystem, this will cause all data to be erased

*hadoop@n1:/usr/local/hadoop$ bin/hadoop namenode –format*

Now run following command which will startup a Namenode, Datanode, Jobtracker and a Tasktracker on the machine.

*hadoop@n1:/usr/local/hadoop$ bin/start-all.sh*

Now run the Java process status tool jps to list all processes in Hadoop.
*hadoop@n1:/usr/local/hadoop$ jps*
5621 JobTracker
5782 TaskTracker
5861 Jps
5545 SecondaryNameNode
5372 DataNode
5200 NameNode

'bin/hadoop dfsadmin' command supports a few HDFS administration related operations.
-report : reports basic stats of HDFS[32]

*hadoop@n1:/usr/local/hadoop$ bin/hadoop dfsadmin –report*

Configured Capacity: 302827593728 (282.03 GB)
Present Capacity: 283720749071 (264.24 GB)
DFS Remaining: 283720724480 (264.24 GB)
DFS Used: 24591 (24.01 KB)
DFS Used%: 0%
Under replicated blocks: 0
Blocks with corrupt replicas: 0
Missing blocks: 0

-------------------------------------------------

Datanodes available: 1 (1 total, 0 dead)
Name: 192.168.0.216:50010
Decommission Status : Normal
Configured Capacity: 302827593728 (282.03 GB)
DFS Used: 24591 (24.01 KB)
Non DFS Used: 19106844657 (17.79 GB)
DFS Remaining: 283720724480(264.24 GB)
DFS Used%: 0%
DFS Remaining%: 93.69%
Last contact: Thu Nov 08 10:09:23 IST 2012

*hadoop@n1:/usr/local/hadoop$    bin/hadoop  dfs  -mkdir datain*

lsr[33] Usage: hadoop fs -lsr <args>
Recursive version of ls. Similar to Unix ls -R.
ls -R : lists directory tree recursively.

*hadoop@n1:/usr/local/hadoop$ bin/hadoop dfs -lsr*
*drwxr-xr-x   - hadoop supergroup          0 2012-11-08 10:12 /user/hadoop/datain*

*hadoop@n1:/usr/local/hadoop$ vi data1.txt*

*Full  virtualization  provides  a  complete  simulation  of underlying  computer  hardware,  enabling  software  to  run without any modification. Because it helps maximize the use and flexibility of computing resources, multiple operating systems can run simultaneously on the same hardware, full virtualization  is  considered  a  key  technology  for  cloud computing. For cloud computing systems, full virtualization can increase operational efficiency because it can optimize computer workloads and adjust the number of servers in use to match demand, thereby conserving energy and information technology resources*

Above data is taken from [36]
*hadoop@n1:/usr/local/hadoop$ vi data2.txt*
*data2.txt also contains same data1.txt data.*

```
hadoop@n1:/usr/local/hadoop$ vi data1.txt
hadoop@n1:/usr/local/hadoop$ vi data2.txt
hadoop@n1:/usr/local/hadoop$ bin/hadoop dfs -put data1.txt datain/
hadoop@n1:/usr/local/hadoop$ bin/hadoop dfs -put data2.txt datain/
hadoop@n1:/usr/local/hadoop$ bin/hadoop jar hadoop-0.20.2-examples.jar wordcount datain dataout
12/11/08 10:27:47 INFO input.FileInputFormat: Total input paths to process : 2
12/11/08 10:27:47 INFO mapred.JobClient: Running job: job_201211081007_0001
12/11/08 10:27:48 INFO mapred.JobClient:  map 0% reduce 0%
12/11/08 10:27:55 INFO mapred.JobClient:  map 100% reduce 0%
12/11/08 10:28:07 INFO mapred.JobClient:  map 100% reduce 100%
12/11/08 10:28:09 INFO mapred.JobClient: Job complete: job_201211081007_0001
12/11/08 10:28:09 INFO mapred.JobClient: Counters: 17
12/11/08 10:28:09 INFO mapred.JobClient:   Job Counters
12/11/08 10:28:09 INFO mapred.JobClient:     Launched reduce tasks=1
12/11/08 10:28:09 INFO mapred.JobClient:     Launched map tasks=2
12/11/08 10:28:09 INFO mapred.JobClient:     Data-local map tasks=2
12/11/08 10:28:09 INFO mapred.JobClient:   FileSystemCounters
12/11/08 10:28:09 INFO mapred.JobClient:     FILE_BYTES_READ=1686
12/11/08 10:28:09 INFO mapred.JobClient:     HDFS_BYTES_READ=1204
12/11/08 10:28:09 INFO mapred.JobClient:     FILE_BYTES_WRITTEN=3442
12/11/08 10:28:09 INFO mapred.JobClient:     HDFS_BYTES_WRITTEN=596
12/11/08 10:28:09 INFO mapred.JobClient:   Map-Reduce Framework
12/11/08 10:28:09 INFO mapred.JobClient:     Reduce input groups=61
12/11/08 10:28:09 INFO mapred.JobClient:     Combine output records=122
12/11/08 10:28:09 INFO mapred.JobClient:     Map input records=22
12/11/08 10:28:09 INFO mapred.JobClient:     Reduce shuffle bytes=1692
12/11/08 10:28:09 INFO mapred.JobClient:     Reduce output records=61
12/11/08 10:28:09 INFO mapred.JobClient:     Spilled Records=244
12/11/08 10:28:09 INFO mapred.JobClient:     Map output bytes=1858
12/11/08 10:28:09 INFO mapred.JobClient:     Combine input records=164
12/11/08 10:28:09 INFO mapred.JobClient:     Map output records=164
12/11/08 10:28:09 INFO mapred.JobClient:     Reduce input records=122
hadoop@n1:/usr/local/hadoop$
```
Fig 5: Screenshot of Word-count Map-reduce Job

*hadoop@n1:/usr/local/hadoop$    bin/hadoop  dfs    -cat /user/hadoop/output1/part-r-00000*

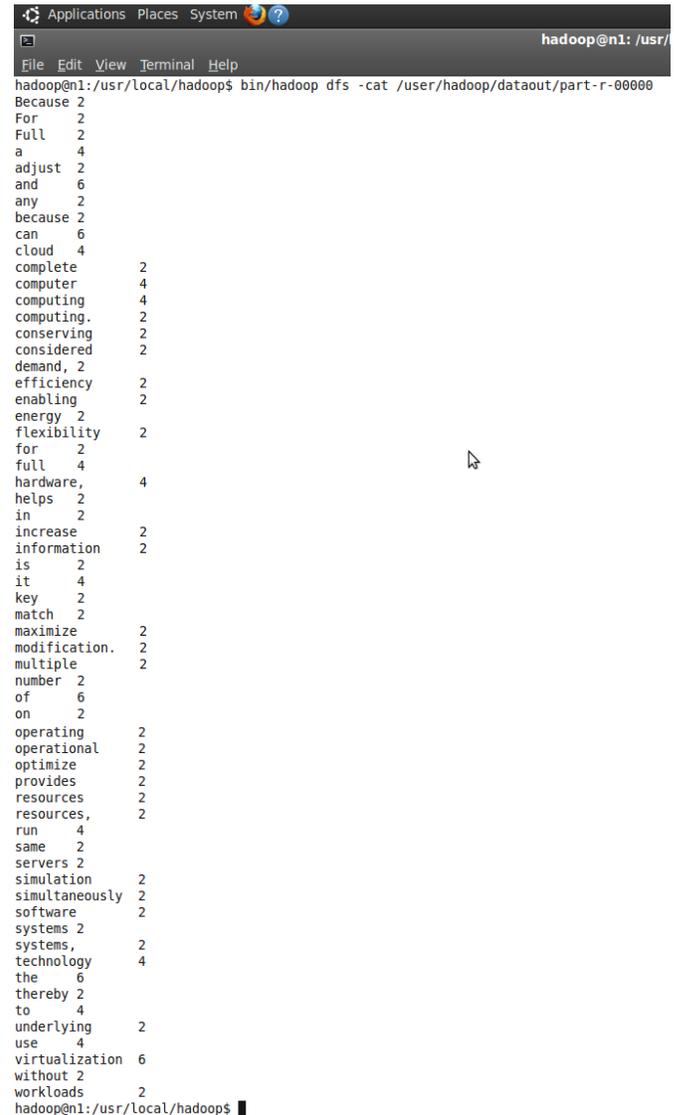

Fig 6: Screenshot of output of Wordcount Map-reduce Job

### NameNode Web Interface (HDFS layer)

The name node web UI depicts a cluster summary including information about total/remaining capacity, live and dead nodes. Additionally, it allows user to browse the HDFS namespace and view the contents of its files in the web browser. It also gives access to the local machine's Hadoop log files [6].

By default, it's available at http://localhost:50070/.

## NameNode 'n1:9000'

| | |
|---|---|
| Started: | Thu Nov 08 10:07:36 IST 2012 |
| Version: | 0.20.2, r911707 |
| Compiled: | Fri Feb 19 08:07:34 UTC 2010 by chrisdo |
| Upgrades: | There are no upgrades in progress. |

**Browse the filesystem**
**Namenode Logs**

### Cluster Summary

**21 files and directories, 10 blocks = 31 total. Heap Size is 59.25 MB / 888.94 MB (6%)**

| | | |
|---|---|---|
| Configured Capacity | : | 282.03 GB |
| DFS Used | : | 108 KB |
| Non DFS Used | : | 17.82 GB |
| DFS Remaining | : | 264.21 GB |
| DFS Used% | : | 0 % |
| DFS Remaining% | : | 93.68 % |
| **Live Nodes** | : | 1 |
| **Dead Nodes** | : | 0 |

### NameNode Storage:

| Storage Directory | Type | State |
|---|---|---|
| /home/hadoop/cloud/hadoop-hadoop/dfs/name | IMAGE_AND_EDITS | Active |

**Hadoop**, 2012.

Fig 7: Screenshot of NameNode

### JobTracker Web Interface (MapReduce layer)

The job tracker web UI provides information about general job statistics of the Hadoop cluster, running/completed/killed jobs and a job history log file. It also gives access to the local machine's Hadoop log files (the machine on which the web UI is running on)[6].
By default, it's available at http://localhost:50030/.

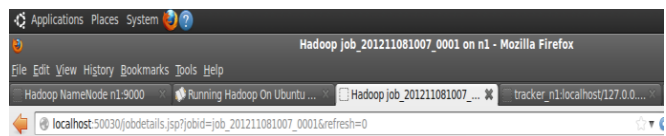

Fig 8: Screenshot of JobTracker

### TaskTracker Web Interface (MapReduce layer)

The task tracker web UI depicts running and non-running tasks. It also gives access to the local machine's Hadoop log files [6].
By default, it's available at http://localhost:50060/.

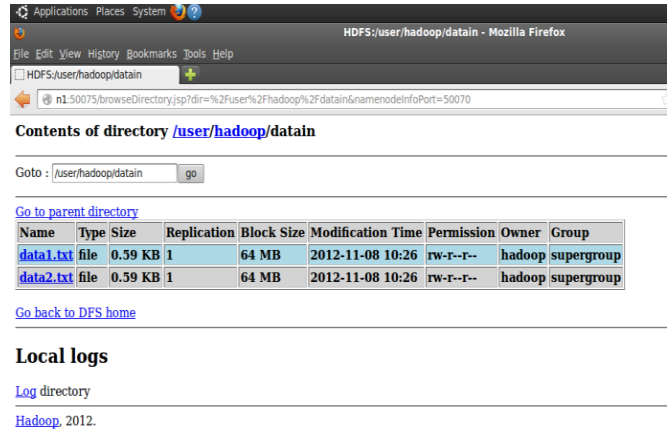

Fig 9: Screenshot of text files in Hadoop

The two data text files are reduced to one text file saving 64 MB of block size using Hadoop. Combined two files contained 164 words which is map-reduced to one file of 61 records.

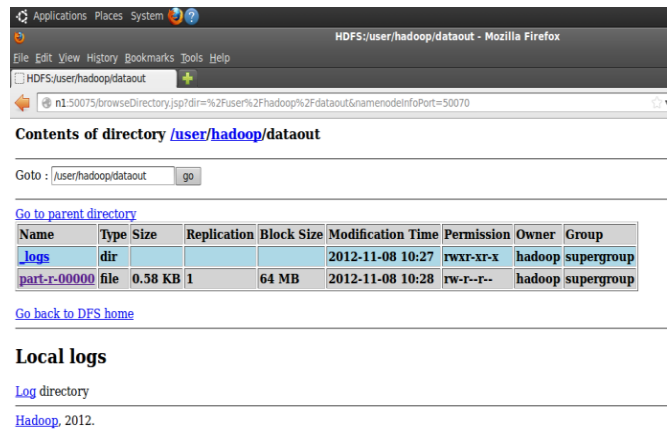

Fig 10: Screenshot of output file after Map-reduce

## 4.0 HBase Installation

Download hbase-0.94.2.tar.gz or higher version of HBase from Apache Download Mirrors. Extract it into a folder and change conf/hbase-site.xml file, set hbase.rootdir which is the directory where HBase writes data to, and hbase.zookeeper.property.dataDir, the directory ZooKeeper writes its data too [25]. ZooKeeper is a centralized service for maintaining configuration information, naming, providing distributed synchronization, and providing group services [24].

```
<property>
    <name>hbase.rootdir</name>
    <value>hdfs://master:9000/hbase</value>
</property>
<property>
    <name>hbase.zookeeper.property.dataDir</name>
    <value>/home/nandan/usr/local/hadoop/hbase/hbase-0.90.5/tmp</value>
</property>
```

Fig 11: Screenshot of hbase-site.xml

Now start the HBase using following command

*$ ./bin/start-hbase.sh*

```
hadoop@n1:/usr/local/hadoop/hbase/hbase-0.90.5$ bin/start-hbase.sh
localhost: starting zookeeper, logging to /usr/local/hadoop/hbase/hbase-0.90.5/bin/../logs/hbase-hadoop-zookeeper-n1.out
starting master, logging to /usr/local/hadoop/hbase/hbase-0.90.5/bin/../logs/hbase-hadoop-master-n1.out
n1: starting regionserver, logging to /usr/local/hadoop/hbase/hbase-0.90.5/bin/../logs/hbase-hadoop-regionserver-n1.out
hadoop@n1:/usr/local/hadoop/hbase/hbase-0.90.5$ jps
5402 Jps
2218 JobTracker
1954 DataNode
5121 HQuorumPeer
2302 TaskTracker
5171 HMaster
5342 HRegionServer
1790 NameNode
2123 SecondaryNameNode
hadoop@n1:/usr/local/hadoop/hbase/hbase-0.90.5$
```
Fig 12: Screenshot of starting HBase

Now connect to running Hbase via shell as follows

*hadoop@n1:/usr/local/hadoop/hbase/hbase-0.90.5$*
*./bin/hbase shell*

Using shell commands a sample table is created in Hbase named 'student' which has an attribute 'name'. In this table compression is not specified.

```
hbase(main):001:0> create 'student', 'name'
0 row(s) in 1.6420 seconds

hbase(main):002:0> put 'student', 'row1', 'name:a', 'Nandan'
0 row(s) in 0.1390 seconds

hbase(main):003:0> put 'student', 'row2', 'name:b', 'Sandeep'
0 row(s) in 0.0060 seconds

hbase(main):004:0> put 'student', 'row3', 'name:c', 'Aaradhana'
0 row(s) in 0.0070 seconds

hbase(main):005:0> scan 'student'
ROW                    COLUMN+CELL
 row1                  column=name:a, timestamp=1350023991906, value=Nandan
 row2                  column=name:b, timestamp=1350024019287, value=Sandeep
 row3                  column=name:c, timestamp=1350024031339, value=Aaradhana
3 row(s) in 0.0440 seconds
```
Fig 13: Screenshot of running HBase and Shell Commands

Using exit command one can come out of running Hbase.
To stop HBase use following command

*$ ./bin/stop-hbase.sh*

# 5.0 LZO Compression

In order to allow compression in HBase itself a compression software has to be installed in Hadoop. In this experiment LZO compression is used to compress the data which is as follows.

We will use the hadoop-lzo library to add LZO compression support to HBase: [26]

**1**. Get the latest hadoop-lzo source from https://github.com/toddlipcon/hadoop-lzo.
**2**. Build the native and Java hadoop-lzo libraries from source. Depending on your OS, to build 64-bit binaries run the following commands:
$ export CFLAGS="-m64"
$ export CXXFLAGS="-m64"
$ cd hadoop-lzo
$ ant compile-native
$ ant jar
These commands will create the hadoop-lzo/build/native directory and the hadoop-lzo/build/hadoop-lzo-x.y.z.jar file. In order to build 32-bit binaries, just change the value of CFLAGS and CXXFLAGS to –m32.

**3**. Copy the built libraries to the $ hbase-0.94.2/lib and $ hbase-0.94.2/lib/native directories on node:

$ cp hadoop-lzo/build/hadoop-lzo-x.y.z.jar hbase-0.94.2/lib
$ mkdir hbase-0.94.2/lib/native/Linux-amd64-64
$ cp hadoop-lzo/build/native/Linux-amd64-64/lib/* hbase-0.94.2/lib/native/Linux-amd64-64

**4**. Add the configuration of hbase.regionserver.codecs to your hbase-site.xml file:

$ vi hbase-0.94.2/conf/hbase-site.xml

<property>
<name>hbase.regionserver.codecs</name>
<value>lzo,gz</value>
</property>

**5**. Sync the $ hbase-0.94.2/conf and $ hbase-0.94.2/lib directories across the cluster.
**6**. HBase ships with a tool to test whether compression is set up properly. Use this tool to test the LZO setup. If everything is configured accurately, we get the SUCCESS output:

```
root@n1:/usr/local/hadoop/hbase-0.94.2# ./bin/hbase org.apache.hadoop.hbase.util.CompressionTest file:///usr/local/hadoop/data2.txt lzo
12/11/27 16:20:46 INFO util.ChecksumType: org.apache.hadoop.util.PureJavaCrc32 not available.
12/11/27 16:20:46 INFO util.ChecksumType: Checksum can use java.util.zip.CRC32
12/11/27 16:20:46 INFO util.ChecksumType: org.apache.hadoop.util.PureJavaCrc32C not available.
12/11/27 16:20:46 DEBUG util.FSUtils: Creating file:file:/usr/local/hadoop/data2.txtwith permission:rwxrwxrwx
12/11/27 16:20:46 INFO lzo.GPLNativeCodeLoader: Loaded native gpl library
12/11/27 16:20:46 INFO lzo.LzoCodec: Successfully loaded & initialized native-lzo library
12/11/27 16:20:46 INFO compress.CodecPool: Got brand-new compressor
12/11/27 16:20:46 DEBUG hfile.HFileWriterV2: Initialized with CacheConfig:disabled
12/11/27 16:20:46 INFO compress.CodecPool: Got brand-new decompressor
SUCCESS
root@n1:/usr/local/hadoop/hbase-0.94.2#
```
Fig 14: Screenshot of Success of Loading native-lzo/native gpl library

7. Test the configuration by creating a table with LZO compression.

```
hbase(main):001:0> create 'Faculty', {NAME=>'name', COMPRESSION=>'lzo'}
0 row(s) in 1.9500 seconds

hbase(main):002:0> put 'Faculty', 'row1', 'name:a', 'Nandan'
0 row(s) in 0.1010 seconds

hbase(main):003:0> put 'Faculty', 'row2', 'name:b', 'Sandeep'
0 row(s) in 0.0140 seconds

hbase(main):004:0> put 'Faculty', 'row3', 'name:c', 'Aaradhana'
0 row(s) in 0.0090 seconds

hbase(main):005:0> scan 'Faculty'
ROW                     COLUMN+CELL
 row1                   column=name:a, timestamp=1354014818509, value=Nandan
 row2                   column=name:b, timestamp=1354014827813, value=Sandeep
 row3                   column=name:c, timestamp=1354014836738, value=Aaradhana
3 row(s) in 0.0500 seconds
```

Fig 15: Screenshot of creating table in HBase with LZO compression

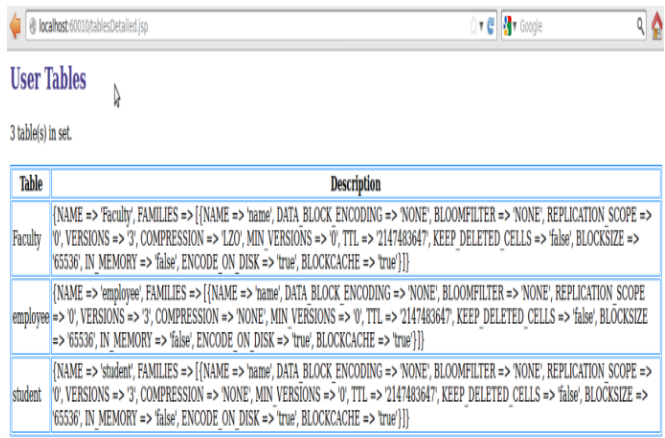

Fig 16: Screenshot of Tables in HBase

By adding LZO compression support, HBase StoreFiles (HFiles) will use LZO compression on blocks as they are written. HBase uses the native LZO library to perform the compression, while the native library is loaded by HBase via the hadoop-lzo Java library that is built. In order to avoid starting a node with any codec missing or misinstalled, add LZO to the hbase.regionserver.codecs setting in the hbase-site.xml file. This setting will cause a failed startup of a region server if LZO is not installed properly. "Could not load native gpl library" is visible then there is an issue with the LZO installation. In order to fix it, the native LZO libraries are installed and the path is configured properly. A compression algorithm is specified on a per-column family basis. We create a table Faculty, with a single column family name, which uses LZO compression on it. Although it adds a read-time penalty as the data blocks probably a decompressed when reading, LZO is fast enough as a real-time compression library. We recommend using LZO as the default compression algorithm in production HBase [26].

# 6.0 Conclusion

The LZO compression format was designed considering speed as priority, it decompresses about twice as fast as gzip, meaning it's fast enough to keep up with hard drive read speeds. It doesn't compress quite as well as gzip— expect files that are on the order of 50% larger than their gzipped version. But that is still 20-50% of the size of the files without any compression at all, which means that IO-bound jobs complete the map phase about four times faster [34]. Following table is a typical example, starting with an 8.0 GB file containing some text-based log data:

| Compression | File | Size (GB) | Compression Time (s) | Decompression Time (s) |
|---|---|---|---|---|
| None | some_logs | 8.0 | - | - |
| Gzip | some_logs.gz | 1.3 | 241 | 72 |
| LZO | some_logs.lzo | 2.0 | 55 | 35 |

Fig 17: Comparison of different compression formats [34]

As per above table the LZO file is slightly larger than the corresponding gzip file, but both are much smaller than the original uncompressed file. Additionally, the LZO file compressed nearly five times faster, and decompressed over two times faster [34].

In this way paper illustrates the importance of Hadoop in current Big-Data world, power of Map-reduce algorithm and necessity of compression of data.

Scottsdale, Arizona, USA. Copyright 2012 ACM 978-1-4503-1247-9/12/05 ...$10.00.

**First Author**: **Nandan Nagarajappa Mirajkar** is pursuing **M.Tech** in Advanced Information Technology with specialization in Software Technologies from IGNOU – I$^2$IT Centre of Excellence for Advanced Education and Research, Pune, India. He is also **Teaching Assistant** in Advanced Software and Computing Technologies department. He has published one International Journal in IJCSI. He is IGNOU – I$^2$IT Centre of Excellence for Advanced Education and Research Academic Scholarship holder. He has pursued **B.E** Electronics and Telecommunications from University of Mumbai. His research interests include Cloud computing, Databases and Networking.

**Second Author**: **Sandeep Bhujbal** is **Sr. Research Associate** in Advanced Software and Computing Technologies department of IGNOU – I$^2$IT Centre of Excellence for Advanced Education and Research, Pune, India. He has pursued **M.C.S** from University of Pune. His research interests include Operating systems, Compiler construction, Programming languages and Cloud computing.

**Third Author**: **Prof. (Ms.) Aaradhana Arvind Deshmukh** is Asst.Professor in Dept. of Computer Engineering , Smt. Kashibai Navale Collge of Engineering, Pune. She is **pursuing PhD in Cloud Computing from Aalborg University, Denmark**. She is visiting Faculty in Advanced Software and Computing Technologies department of IGNOU – I$^2$IT Centre of Excellence for Advanced Education and Research, Pune, India. She obtained Masters [Computer Engineering ] , A.M.I.E. Computer Engineering , B.E. (Computer Engineering) M.A. (Economics) from Pune University. She is having 10 years experience in Teaching Profession and 2 ½ years R & D experience in various institutes under Pune University. She has published 43 papers , 13 in International Journals like ACM, IJCSI, ICFCA, IJCA etc, 16 in International Conferences like IEEE etc., 9 in National Conferences, 4 in symposiums . She has received Gold Medal at International level Paper Presentation on "Neural Network" as well as one more for " UWB Technology based adhoc network", in International Conferences. She is recipient of 'Distinguished Alumni Award' in 2011 from Inst. Of Engineers [India] , Gunawant Nagrik Puraskar for the year 2004 – 2005, 'Anushka Purskar' from Pimpri Chinchwad Municipal Corporation, and also won many Firodiya awards. She has organized many 15 multidisciplinary Short Term Training Program, workshops, conferences on National and International Level. Her research interests include Cloud computing, Databases and Networking, security.